\documentclass[reprint,superscriptaddress,showpacs,nofootinbib,amsmath,amssymb,aps,prl]{revtex4-1}
\usepackage{graphicx}% Include figure files
\usepackage{dcolumn}% Align table columns on decimal point
\usepackage{bm}% bold math
\usepackage[utf8]{inputenc}
\usepackage[frenchb]{babel}
\usepackage{braket}

\begin{document}

\preprint{APS/123-QED}

\title{Autolocalization in a dipolar exciton system}
\author{S. V. Andreev}
\email[Electronic adress: ]{Serguei.Andreev@u-psud.fr}
\affiliation{CNRS, LPTMS, Université Paris Sud, UMR8626, 91405 Orsay, France}
\affiliation{ITMO University, St. Petersburg 197101, Russia}

\date{\today}

\begin{abstract}
We develop the autolocalization hypothesis suggested recently in [Andreev, Phys. Rev. Lett. \textbf{110}, 146401 (2013)] to explain the formation of the macroscopically ordered exciton state (MOES) in semiconductor quantum wells [L.~V.~Butov \textit{et al.}, Nature
(London) \textbf{418}, 751 (2002)]. We argue that the onset of a periodical localizing potential having a macroscopic spatial period is possible in the systems where in addition to long-range dipolar repulsion the excitons exhibit resonant pairing at short distances. Our theory suggests, that the central incoherent part of each condensate in the MOES may represent a novel quantum molecular phase, which was predicted and discussed theoretically several years ago in the context of resonant Bose superfluids.   
\end{abstract}


\maketitle

It is well known that there is no Bose-Einstein condensation (BEC) in 1D and 2D bosonic gases at a finite temperature, because the off-diagonal long-range order (ODLRO) is destroyed by long-wave phase fluctuations \cite{Hohenberg, Pitaevskii}. To observe the true second-order phase transition in a disorder-free low-dimensional system one has to introduce an external trapping potential. Localization supresses the phase fluctuations \cite{Petrov1, Petrov2} and brings about criticality in thermodynamics by transforming the density of single-particle states (DOS) \cite{thermodynamics, TrappedGas}.

Following these arguments, the author has recently postulated that the macroscopic ordering of excitons discovered \cite{Butov2002} and recently duplicated  \cite{Alloing} in high-quality semiconductor quantum wells (QW's) can be considered as a manifestation of \textit{autolocalization} \cite{Andreev1}. Below few degrees Kelvin a uniform quasi-one-dimensional (quasi-1D) gas of dipolar excitons generated in the ring-shaped trap transforms into a chain of macroscopic aggregates seen as bright spots in the exciton photoluminescense (PL) pattern. Temperature dependence of the exciton energy exhibits the typical casp at the transition point \cite{Andreev1, Repulsive}. Each aggregate consists of a bright core surrounded by a coherent halo of weaker PL intensity. Remarkably, the coherence of the PL in the halo is extended over the length scale comparable with the size of an aggregate \cite{High2012}. These experimental facts have been phenomenologically taken into account in \cite{Andreev1} by treating the aggregates as trapped Bose-Einstein condensates, the trapping along the chain being the result of electrostatic repulsion between the neighbors. The model allows one to estimate the number of condensates at the ring in equilibrium and to explicitly demonstrate the scale invariance and universality of the phenomenon \cite{Andreev2} which are known to be the distinct features of the second-order phase transition \cite{Kadanoff}.

Until now, however, microscopic origin of the autolocalization has not been understood. In particular, it has been unclear why the dramatic phenomena observed by Butov group \cite{Butov2002, High2012} do not take place in the experimental configuration employed by Snoke \cite{SnokeNature} which corresponds essentially to the same set of parameters.

In this Letter we describe the onset of the autolocalizing potential along the ring by considering quantum scattering of dipolar excitons in the fluctuation region near the phase transition. We show that coherent crystallization of the exciton gas may occur in the QW structures where, in addition to long-range dipolar repulsion, the excitons exhibit \textit{resonant pairing} at short distances. A resonance in the exciton-exciton interaction potential may appear when the distance between electron and hole layers $d$ approaches (from above) the critical value $d_c$ which corresponds to the biexciton dissociation threshold estimated in \cite{Zimmermann, Fogler}. Numerical studies \cite{Zimmermann} indicate, that this unique situation may be realized in the sample used in \cite{Butov2002, High2012}. In such sample beyond the fluctuation region Bose-Einstein condensates are formed in each site of the autolocalizing potential. Strong electrostatic repulsion freezes the tunneling between the neighbours thus stabilizing the condensates againt the long-wave collective excitations. Mean-field analysis shows that each condensate consists of two distinct excitonic and molecular phases. The molecular phase occupies the central dense part of a condensate and is distinguished by the absence of excitonic ODLRO.

According to Landau theory of second-order phase transitions \cite{Landau}, Bose-Einstein condensation (BEC) of excitons localized at the ring should be heralded by coherent fluctuations of the exciton order parameter above the corresponding critical temperature $T_c$. As one gets close to $T_c$, these fluctuations cover a large distance $L$ in the azimuthal direction. The probability to find the excitons in a coherent many-body state $\ket\Phi$ at fixed $T$ and chemical potential $\mu(T)$ can be evaluated as \cite{Landau}
\begin{equation}
\label{w}
w_\Phi\sim \exp\left(-\bra{\Phi}\hat\Omega\ket{\Phi}/k_B T\right),
\end{equation} 
where $k_B$ is the Boltzmann constant and the fluctuation part $\hat\Omega$ of the grand canonical Hamiltonian of the system reads
\begin{equation}
\label{SpinlessHamiltonian}
\begin{split}
\hat\Omega=&\int\hat\Psi^\dagger(\bm \rho)\left(-\frac{\hbar^2}{2m}\Delta+\frac{1}{2}m\omega_y^2 y^2-\mu\right)\hat\Psi(\bm \rho) d\bm \rho\\
&+\frac{1}{2}\int\hat\Psi^\dagger(\bm \rho)\hat\Psi^\dagger(\bm \rho')V(\bm \rho-\bm \rho')\hat\Psi(\bm \rho')\hat\Psi(\bm \rho)d\bm \rho d\bm \rho',
\end{split}
\end{equation}
with $\bm \rho=(x,y)$ being the coordinate of exciton translational motion in the structure plane. The exciton field operator  $\hat\Psi (\bm\rho)$ can be conveniently recast in the form
\begin{equation}
\label{Psi}
\hat\Psi(x,y)=\phi(y)\hat\psi(x),
\end{equation}
where we assume tight harmonic confinement in the radial direction $y$,
\begin{equation}
\phi(y)=\frac{1}{(\sqrt{\pi}a_y)^{1/2}}e^{-y^2/2a_y^2},
\end{equation}
with $a_y=\sqrt{\hbar/m\omega_y}$ being the corresponding oscillator length, and impose cyclic boundary conditions along the ring
\begin{equation}
\hat\psi(x)=\frac{1}{\sqrt{L}}\sum_{k_x} \hat c_{k_x}e^{ik_x x},
\end{equation}
with $k_x=\{0,\pm 2\pi/L,\pm 4\pi/L,...\}$. At the present stage we neglect the spinor nature of $\hat\Psi$. Modification of the theory to account for the exciton spin will be discussed later.

The presence of a gradient term in the expansion \eqref{SpinlessHamiltonian} suggests that, according to \eqref{w}, the fluctuation modes which vary slowly along the ring will have the largest probability $w_\Phi$. In particular, for a system interacting via an (effective) short-range potential
\begin{equation}
\label{cont}
V(\bm \rho-\bm \rho')=V_0 \delta(\bm \rho-\bm \rho'),
\end{equation}
with $V_0>0$ (repulsion), an \textit{optimal fluctuation} would be
\begin{equation}
\label{Uniform}
\ket{\Phi_0}=\frac{1}{\sqrt{N!}}(\hat c_0^\dagger)^N\ket{\mathrm{vac}},
\end{equation}
where all $N\gg 1$ fluctuating excitons occupy the single-particle state with $k_x=0$.
In what follows we shall assume $k_x a_y\ll 1$ and explain, how this simple picture should be revised when taking into account the dipolar tail of the two-body potential 
\begin{equation}
\label{tail}
V(\bm \rho-\bm \rho')\approx V_\ast(x-x')=\frac{\hbar^2}{m}\frac{x_\ast}{\lvert x-x'\rvert^3},
\end{equation}
where $\lvert x-x'\rvert\gg a_y\gtrsim x_\ast$ and $x_\ast=m\mathrm{e}^2d^2/4\pi\hbar^2\epsilon\epsilon_0$ is the characteristic dipole-dipole distance.
  
To work out the interaction of the fluctuation modes with small momenta in the simplest form, we take advantage of the fact that in the critical region $T\rightarrow T_c$ the system is dilute and one can follow the Beliaev prescription \cite{Beliaev, Abrikosov} to replace the actual microscopic potential $V(\bm \rho-\bm \rho')$ by the corresponding $s$-wave scattering amplitude. This replacement can be done in two steps. We first substitute \eqref{Psi} and \eqref{cont} into the Hamiltonian \eqref{SpinlessHamiltonian} to obtain a quasi-1D coupling constant $V_0^{1D}=V_0/\sqrt{2\pi}a_y$ describing the contact part of the interaction. Next, we include perturbatively the contribution coming from distances of the order of $k_x^{-1}$, where the relative motion of excitons becomes one-dimensional and is governed by the dipolar tail \eqref{tail}. This way we obtain \cite{SI}
\begin{widetext}
\begin{equation}
\label{Hamiltonian}
\hat\Omega_\ast=\sum_{k_x}\frac{\hbar^2 k_x^2}{2m}\hat c^\dagger_{k_x}\hat c_{k_x}+
\frac{V_0^{1D}}{2L}\sum_{k_x, p_x, q_x}[1+2e\zeta(\lvert p_x-q_x\rvert x_\ast)^2\ln(\lvert p_x-q_x\rvert x_\ast)]\hat c^\dagger_{k_x+p_x}\hat c^\dagger_{k_x-p_x}\hat c_{k_x+q_x}\hat c_{k_x-q_x},
\end{equation}
\end{widetext}
where $\hat c^\dagger_{k_x}$ and $\hat c_{k_x}$ are the fluctuation mode creation and annihilation operators, the dimensionless parameter $\zeta$ is defined by
\begin{equation}
\label{zeta}
\zeta=\frac{\sqrt{2\pi}}{2e}\frac{\hbar^2}{mV_0}\frac{a_y}{x_\ast},
\end{equation}
with the prefactor $2e$ ($e$=2.718...) being introduced for future convenience and we have taken into account that $\mu(T\rightarrow T_c)\approx\hbar\omega_y/2$.

The logarithmic momentum-dependent term in the ersatz Hamiltonian \eqref{Hamiltonian} describes the effect of long-range dipolar interaction on the fluctuating excitons and it alters dramatically their collective behaviour as compared to the result \eqref{Uniform}. Namely, it can be shown \cite{SI}, that as the one-dimensional density $n_1\equiv N/L$ of a fluctuation becomes larger than 
\begin{equation}
\label{n_c}
n_c=(x_\ast \ln\zeta)^{-1},
\end{equation}
the expectation value $\bra{\Phi}\hat\Omega_\ast\ket{\Phi}$ is minimized not by the spatially uniform configuration \eqref{Uniform}, but by
\begin{equation}
\label{CC}
\ket{\Phi_k}=\frac{1}{\sqrt{N!2^N}}\left(\hat c_{-k}^\dagger+\hat c_{k}^\dagger\right)^N\ket{\mathrm{vac}},
\end{equation}
where for $n_1=n_c$ the wave-vector $k$ is given by
\begin{equation}
\label{k_c}   
k=k_c\approx0.14(\sqrt{\zeta} x_\ast)^{-1}.
\end{equation}
The coherent many-body state \eqref{CC} represents a crystalline structure where a periodic self-consistent field
\begin{equation}
\label{Auto}
\begin{split}
V_\mathrm{auto}(x)\equiv&\bra{\Phi_{k}}\int\hat\psi^{\dagger}(x')V_\ast(x-x')\hat\psi(x')dx'\ket{\Phi_{k}}\\
&=-\frac{\hbar^2n_1}{2m x_\ast^2}(2kx_\ast)^2\ln(2kx_\ast)\cos(2k x)
\end{split}
\end{equation}
and a periodic density distribution 
\begin{equation}
\label{Density}
n_1(x)\equiv\bra{\Phi_{k}}\hat\psi^{\dagger}(x)\hat\psi(x)\ket{\Phi_{k}}=n_1\cos^2(k x)
\end{equation}
maintain each other.

A possibility for the system to produce a fluctuation of the type \eqref{CC} requires the magnitude of the autolocalizing potential \eqref{Auto} to be of the same order as an increase of the contact interaction energy due to the exciton density modulation \eqref{Density}. This limits the relevant parameter to
\begin{equation}
\label{Cond}
\zeta\gg 1.
\end{equation}
In the opposite limit, when $\zeta\rightarrow 1$, the activation energy 
\begin{equation}
\label{Energy}
\Omega_c\equiv\bra{\Phi_{k}}\hat\Omega_\ast\ket{\Phi_{k}}\lvert_{k=k_c}=\frac{\hbar^2}{mx_{\ast}^2}\frac{2e}{\zeta\ln\zeta}
\end{equation}
becomes increasingly large, with the consequence that the system will remain in the uniform state \eqref{Uniform}. Clearly, the latter holds for $\zeta\leqslant 1$ as well.  

At a first glance, the requirement \eqref{Cond} is never fulfilled in real systems, which manifest tremendous blueshift of the exciton PL line below $T_c$ \cite{Repulsive, Alloing}. Indeed, for the experiments \cite{Butov2002, Alloing}, where the macroscopic ordering of excitons has been observed, we estimate $\zeta\sim 0.1$, owing to a large value of the 2D coupling constant $V_0$. This observation presents an apparent challenge to our theory.    

To explain the buildup of a macroscopic spatial order in the systems characterized by strong repulsion, one should go beyond the spinless model introduced above. We shall now consider the contact part of the exciton-exciton interaction in more detail, accounting for possible exchange effects between the constituent electrons and holes.

\begin{figure}[t]
\includegraphics[width=0.8\columnwidth]{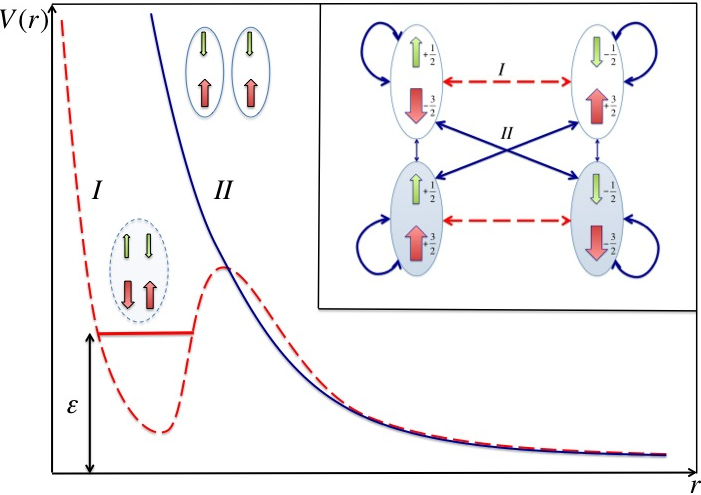}
\caption{Two-body interaction potential as a function of relative distance between the excitons for two types of scattering channels depicted schematically in the inset. The type II interaction is featureless, characterized by strong repulsion at short distances. The type I scattering channel, in which two electrons and two holes in the excitons simultaneously have anti-parallel spins, can admit a quasi-bound state (resonance) with positive energy and finite lifetime.}
\label{Fig1}
\end{figure}

At small interlayer distance $d$ the exchange interaction is known to prevail over the dipolar repulsion, leading to binding of excitons into molecules (biexcitons) in the scattering channel where both electrons and holes in the colliding excitons have anti-parallel spins (this channel is denoted as "type I" in the inset of Fig. \ref{Fig1}). At low temperatures this may result in attractive interaction between the excitons \cite{Sugakov} and favor formation of metallic electron-hole droplets \cite{Butov2007}, which precludes observation of an exciton BEC. As one increases $d$, the biexciton binding energy decreases, until, at some critical value $d_c$ the true bound state disappears \cite{Fogler}.

It is natural to expect, however, that for $d\gtrsim d_c$ the corresponding two-body potential (dashed line in Fig. \ref{Fig1}) can admit a \textit{quasi-bound} state (\textit{resonance}) with positive (with respect to the scattering threshold) energy and finite lifetime. Numerical studies \cite{Zimmermann} indicate, that this unique situation can be realized in the CQW sample used by Butov group \cite{Butov2002, High2012}, suggesting that there is a link between the resonant pairing and coherent crystallization of the exciton ring. In the remainder of this Letter we develop this hypothesis .

We may, without loss of generality, consider only bright or dark excitons \cite{Kavokin}, as the interaction between these two species (denoted as "type II" in Fig. \ref{Fig1}) is featureless and results merely in a positive background shift of the overall interaction energy. Following Refs. \cite{Gurarie, Radzihovsky}, we then adopt the Fano-Anderson model of a discrete level inside a continnum \cite{Fano} to a model system composed of spin up $\hat\Psi_{\uparrow}$, spin down $\hat\Psi_{\downarrow}$ bosons and their bosonic molecules $\hat\Psi_{B}$. The grand canonical Hamiltonian reads
\begin{equation}
\label{SpinHamiltonian}
\begin{split}
&\hat\Omega'=\int\sum_\sigma\hat\Psi^\dagger_\sigma(\bm \rho)\left(-\frac{\hbar^2}{2m_\sigma}\Delta+\frac{1}{2}m\omega_y^2 y^2-\mu_\sigma\right)\hat\Psi_\sigma(\bm \rho) d\bm \rho\\
&+\frac{1}{2}\int\sum_{\sigma,\sigma'}\hat\Psi^{\dag}_{\sigma}(\bm\rho)\hat\Psi^{\dag}_{\sigma'}(\bm \rho')V_{\sigma'\sigma}(\bm \rho-\bm \rho')\hat\Psi_{\sigma}(\bm\rho)\hat\Psi_{\sigma'}(\bm\rho')d\bm \rho' d\bm \rho\\
&-\frac{\alpha}{2}\int(\hat\Psi^\dagger_\uparrow\hat\Psi^\dagger_{\downarrow}\hat\Psi_{B}+\hat\Psi_\uparrow\hat\Psi_{\downarrow}\hat\Psi^\dagger_{B})d\bm \rho+\varepsilon\int\hat\Psi_{B}^\dagger\hat\Psi_{B}d\bm\rho
\end{split}
\end{equation}
where $\sigma=\{\uparrow,\downarrow,B\}$ labels a flavor of the exciton field, $m_{\uparrow}=m_{\downarrow}=m$ and $m_{B}=2m$ being the exciton and quasi-biexciton effective masses correspondingly and the chemical potentials satisfy $\mu_\uparrow=\mu_\downarrow=\mu_B/2\equiv\mu$. The coupling constant $\alpha$  characterizes the coherent exciton-biexciton interconversion rate (hybrydization between the resonance and the outer scattering states induced by tunneling under the potential barrier). The details on the background potentials $V_{\sigma'\sigma}$ can be found in \cite{SI}. The energy of a biexciton at rest $\varepsilon$ depends on the interlayer distance $d$ as detailed above.

Proceeding along the lines of the spinless theory introduced hitherto, one arrives at the results \eqref{n_c}, \eqref{k_c}, \eqref{Auto} and \eqref{Density}, where now the replacement
\begin{equation}
\label{repl} 
V_0\rightarrow V_0+\frac{\alpha^2}{8(2n_1 V_0^{1D}-\varepsilon)}
\end{equation}
in the definition \eqref{zeta} of the parameter $\zeta$ should be done, and the (quasi-)1D density wave $n_1(x)$ becomes equally shared by the two spin components of the exciton field, so that $n_{\uparrow}(x)=n_{\downarrow}(x)=n_1(x)/2$. We have assumed the molecular density being small compared with $n_\uparrow$ and $n_\downarrow$, which is appropriate for the transition point (as before, we assume $T\rightarrow T_c$).

The result \eqref{repl} shows, that resonant pairing of dipolar excitons at short distances may act as an effective attraction, when the energy of a fluctuating exciton pair (given by $2n_1V_0^{1D}$) approaches the resonance from below. This results in a decrease of the activation energy \eqref{Energy}, thus favoring crystallization of the system \cite{Footnote1}. At this stage it might be insightful to complement our static picture with dynamical arguments similar to those given in \cite{Levitov}. A coherent fluctuation \eqref{CC} stimulates scattering of excitons to the single-particle states corresponding to local minima of the autolocalizing potential \eqref{Auto}. In combination with the external confinement along $y$, the latter then may be regarded as a \textit{seed lattice}, into which the excitons start to condense at $T=T_c$, forming a pattern of regularly spaced beads.
\begin{figure}[t]
\includegraphics[width=0.8\columnwidth]{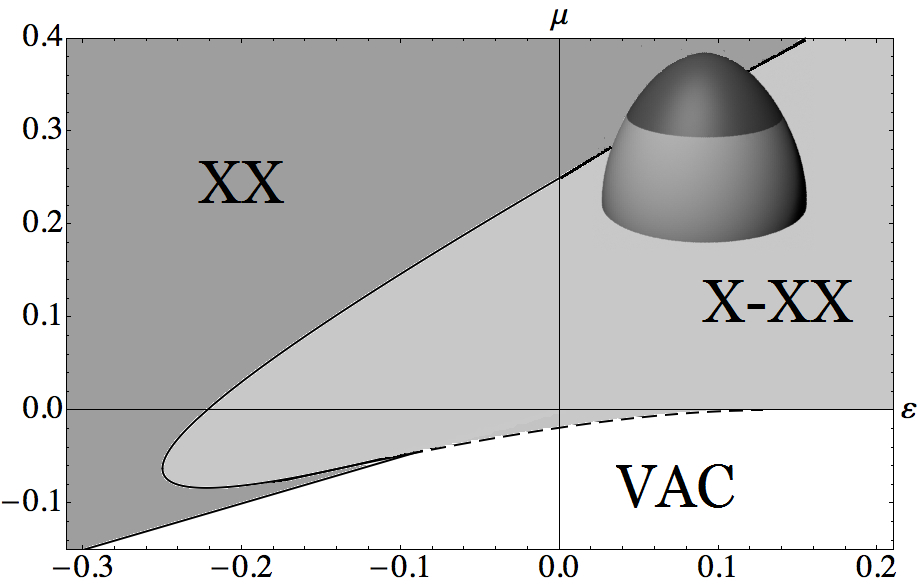}
\caption{Mean-field phase digram of a resonant exciton condensate. VAC is the state vacuum of particles. Light grey area (X-XX) corresponds to the mixture of excitons and their biexcitonic molecules. The dark grey color is used for the pure molecular phase (XX). Solid lines depict continuous phase transition boundaries, whereas the dashed line marks the boundary of the first order \cite{SI}. In the inset we show the LDA energy profile of a harmonically trapped condensate, where one can observe the X-XX/XX transition in real space.}
\label{Fig2}
\end{figure}   

With the substitution $n_1V_0^{1D}\rightarrow\mu$ the effective interaction \eqref{repl} may be employed for condensed excitons as well. However, its validity is now limited to a small peripheric region around each bead, where the system remains dilute. As one moves towards the dense cores, the (local) chemical potential $\mu$ increases and the system exhibits a crossover to a strongly repulsive molecular phase, where all excitons are paired. This behaviour can be better understood from Fig. \ref{Fig2}, where we present a mean-field phase diagram of a uniform 2D system, described by the Hamiltonian \eqref{SpinHamiltonian} with $\omega_y\equiv 0$. In the inset we show a sketch of a 2D harmonically trapped condensate energy profile, which can be obtained by translating the line $\varepsilon=\mathrm{const}$ of the diagram by means of the local density approximation (LDA). The quantum molecular phase (dark grey color) occupies the central dense part of the bead and is distinguished by the absence of excitonic ODLRO (and the associated excitonic superfluidity). This important corollary of our theory is consistent with the experimentally observed patterns of spontaneous coherence of exciton PL \cite{High2012}.

Finally, let us briefly discuss the coherence properties of the system. As is known, BEC does not occur in (quasi-)1D Bose gases because of the enhanced role of the long-wave phase fluctuations in such geometries \cite{1Dgases}. Robustness of the macroscopically ordered exciton state (MOES) to these excitations is ensured by strong quantum fluctuations of the Josephson relative phases, which originate from depleted regions in between the beads \cite{Andreev1, RelativePhases}. These fluctuations drive the coherent state \eqref{CC} into a number squeezed fragmented configuration (Fock state), thus setting an ultraviolet cutoff $k\sim k_c$ for the elementary excitation wave vector. This way the MOES represents a chain of independent, but fully coherent 2D condensates \cite{Footnote2}.

To conclude, we have argued that the autolocalization and formation of the MOES should be observed in the QW structures where, in addition to long-range dipolar repulsion, indirect excitons exhibit resonant pairing at short distances. This resolves the long-standing conundrum of the absence of macroscopic ordering and associated BEC in the exciton system studied in \cite{SnokeNature}: the distance between the two CQW layers here is too large for occurence of the resonance in the exciton-exciton interaction. Our theory suggests, that the central incoherent part of each bead in the MOES may represent a novel quantum molecular phase, which was predicted and discussed theoretically several years ago in the context of resonant Bose superfluids \cite{Radzihovsky}.

Stimulating discussions with D. S. Petrov, G. V. Shlyapnikov, A. A. Varlamov, M. M. Glazov, Y. Y. Kuznetsova and J. R. Leonard are aknowledged. The research leading to these results received funding from the European Research Council (FR7/2007-2013 Grant Agreement No. 341197) and from Ministry of Education and Science of the Russian Federation (The Federal Target Program "5-100").


\begin{thebibliography}{1}

\bibitem{Hohenberg} P. C. Hohenberg, Phys. Rev. \textbf{158}, 383 (1967).

\bibitem{Pitaevskii} L.~Pitaevskii and S.~Stringari, \textit{Bose-Einstein Condensation} (Clarendon Press, Oxford, 2003).

\bibitem{Petrov1} D. S. Petrov, M. Holzmann, and G. V. Shlyapnikov, Phys. Rev. Lett. \textbf{84}, 2551 (2000).

\bibitem{Petrov2} D. S. Petrov, G. V. Shlyapnikov, and J. T. M. Walraven, Phys. Rev. Lett. \textbf{85}, 3745 (2000).

\bibitem{thermodynamics} S. Giorgini, L. P. Pitaevskii and S. Stringari, J. Low
Temp. Phys. \textbf{109}, 309 (1997).

\bibitem{TrappedGas} F. Dalfovo \textit{et al.}, Rev. Mod. Phys. \textbf{71}, 3 (1999).

\bibitem{Butov2002} L.~V.~Butov, A.~C.~Gossard, and D.~S.~Chemla, Nature
(London) \textbf{418}, 751 (2002).

\bibitem{Alloing} M. Alloing, M. Beian, M. Lewenstein, D. Fuster, Y. Gonzalez, L. Gonzalez, R. Combescot, M. Combescot, F. Dubin, Europhys. Lett. \textbf{107}, 10012 (2014).

\bibitem{Andreev1} S. V. Andreev, Phys. Rev. Lett. \textbf{110}, 146401 (2013).

\bibitem{Repulsive} Sen Yang, A.~V. Mintsev, A.~T.~Hammack, L.~V.~Butov, and
A.~C.~Gossard, Phys. Rev. B \textbf{75}, 033311 (2007).

\bibitem{High2012} A.~A.~High, J.~R.~Leonard, A.~T.~Hammack, M.~M.~Fogler,
L.~V.~Butov, A.~V.~Kavokin, K.~L.~Campman, A.~C.~Gossard, Nature (London) 
\textbf{483}, 584 (2012).

\bibitem{Andreev2} S. V. Andreev, A. A. Varlamov and A. V. Kavokin, Phys. Rev. Lett. \textbf{112}, 036401 (2014).

\bibitem{Kadanoff} L. P. Kadanoff \textit{et al.}, Rev. Mod. Phys. \textbf{39}, 395 (1967).

\bibitem{SnokeNature} D. Snoke, S. Denev, Y. Liu, L. Pfeiffer and K. West, Nature (London) \textbf{418}, 754 (2002).

\bibitem{Zimmermann} C. Schindler and R. Zimmermann, Phys. Rev. B \textbf{78}, 045313 (2008).

\bibitem{Fogler} A. D. Meyertholen and M. M. Fogler, Phys. Rev. B \textbf{78}, 235307 (2008).

\bibitem{Landau} L. D. Landau and E. M. Lifshitz, \textit{Statistical Physics} (Pergamon Press, Oxford, 1969).

\bibitem{Beliaev} S. T. Beliaev, Sov. Phys. JETP \textbf{34}, 289 (1958).

\bibitem{Abrikosov} A. A. Abrikosov, L. P. Gorkov, and I. E. Dzialoshinski, \textit{Methods of Quantum Field Theory in Statistical Physics} (Dover, New York, 1975). 

\bibitem{SI} See Supplementary Information.

\bibitem{Repulsive} Sen Yang, A.~V. Mintsev, A.~T.~Hammack, L.~V.~Butov, and
A.~C.~Gossard, Phys. Rev. B \textbf{75}, 033311 (2007).

\bibitem{Sugakov} A. A. Chernyuk and V. I. Sugakov, Phys. Rev. B \textbf{74}, 085303 (2006).

\bibitem{Liu} C. S. Liu, H. G. Luo, W. C. Wu, J. Phys.: Condens. Matter \textbf{18}, 9659 (2006).

\bibitem{Butov2007} L.~V.~Butov, J. Phys.: Condens. Matter \textbf{19},
295202 (2007).

\bibitem{Kavokin} For a detailed discussion of the exciton spin structure see J. R. Leonard, Y. Y. Kuznetsova, Sen Yang, L. V. Butov, T. Ostatnicky, A. Kavokin, and A. C. Gossard, Nano Lett. \textbf{9}, 4204 (2009).

\bibitem{Gurarie} V. Gurarie and L. Radzihovsky, Ann. Phys. \textbf{322}, 2-119 (2007). 

\bibitem{Radzihovsky}  L. Radzihovsky, J. Park, and P. B. Weichman
Phys. Rev. Lett. \textbf{92}, 160402 (2004); L. Radzihovsky, P. B. Weichman and J. I. Park, Ann. Phys. \textbf{323}, 2376-2451 (2008). 

\bibitem{Fano} U. Fano, Phys. Rev. \textbf{124}, 1866 (1961).

\bibitem{Levitov} L. S. Levitov, B. D. Simons and L. V. Butov, Phys. Rev. Lett. \textbf{94}, 176404 (2005).

\bibitem{Footnote1} With further increase of the fluctuation density $n_1$, the system is taken through a strong unitarity scattering limit, corresponding to a divergent scattering amplitude. Interaction of the fluctuation modes here is not correctly accounted for by the substitution \eqref{repl}, as the spin up and spin down excitons start to efficiently convert into their biexcitonic molecules. Analysis of the collective behaviour of the system in this regime is beyond the scope of the present work and will be performed elsewhere.

\bibitem{1Dgases} J. W. Kane and L. P. Kadanoff, Phys. Rev. \textbf{155}, 80 (1967); L. Reatto and G. V. Chester, Phys. Rev. \textbf{155}, 88 (1967); D. S. Petrov, G. V. Shlyapnikov and J. T. M. Walraven, Phys. Rev. Lett. \textbf{87}, 050404 (2001).

\bibitem{RelativePhases} L. P. Pitaevskii and S. Stringari, Phys. Rev. Lett. \textbf{87}, 180402 (2001); A. J. Leggett, Rev. Mod. Phys. \textbf{73}, 307 (2001);

\bibitem{Footnote2} Such coherence regime makes the MOES very different from the conventional supersolids, predicted long time ago \cite{Kirzhnits} and actively studied theoretically nowadays \cite{Supersolids}, but never observed experimentally. In particular, the period of the structure \textit{increases} with increasing the exciton density \cite{Commensurability}. This recent experimental finding favors the theoretical model \cite{Andreev1}, where the number of beads at the ring was determined by minimizing the free energy of the system.

\bibitem{Kirzhnits} D. A. Kirzhnits and Yu. A. Nepomnyashchii, Sov. Phys.
JETP \textbf{32}, 1191 (1971).

\bibitem{Supersolids} A. B. Kuklov, N. V. Prokof’ev, and B. V. Svistunov,
Physics \textbf{4}, 109 (2011); M. Boninsegni and N. V. Prokof’ev,
Rev. Mod. Phys. \textbf{84}, 759 (2012); A. V. Paraskevov and T. V. Khabarova, Physics Letters A 
\textbf{368}, 151 (2007). For roton instability in dipolar systems see, e. g., L. Santos, G. V. Shlyapnikov and M. Lewenstein, Phys. Rev. Lett. \textbf{90}, 250403 (2003); A. Boudjemaa and G. V. Shlyapnikov, Phys. Rev. A \textbf{87}, 025601 (2013); Zhen-Kai Lu, D. S. Petrov and G. V. Shlyapnikov, arXiv: 1409.7737 (2015); A. K. Fedorov, I. L. Kurbakov, Y. E. Shchadilova and Yu. E. Lozovik, Phys. Rev. A \textbf{90}, 043616 (2014);

\bibitem{Commensurability} Sen Yang, L. V. Butov, B. D. Simons, K. L. Campman and A. C. Gossard, arXiv: 1502.02101 (2015).













%\bibitem{Ivanov} A. L. Ivanov, Europhys. Lett. \textbf{59}, 586 (2002).

%\bibitem{Pitaevskii} L.~Pitaevskii and S.~Stringari, \textit{Bose-Einstein
%Condensation} (Clarendon Press, Oxford, 2003).

%\bibitem{Varlamov} A. Larkin and A. Varlamov, \textit{Theory of fluctuations in superconductors} (Oxford University Press, 2005).

%\bibitem{Scaling} S. Giorgini, L. P. Pitaevskii, and S. Stringari, Phys. Rev. Lett. \textbf{78}, 21 (1997).
 
\end{thebibliography}
\end{document}